\documentclass[prl,twocolumn,superscriptaddress]{revtex4}

\usepackage{amsmath,amsfonts,amssymb,amstext}
\usepackage{graphicx}

\def\>{\rangle}
\def\<{\langle}
\def\({\left(}
\def\){\right)}

\begin{document}

\title{Two-Qubit Gates for Resonant Exchange Qubits}

\author{Andrew C. Doherty}
\affiliation{Centre for Engineered Quantum Systems, School of Physics, The University of Sydney, Sydney, NSW 2006, Australia}

\author{Matthew P. Wardrop}
\affiliation{Centre for Engineered Quantum Systems, School of Physics, The University of Sydney, Sydney, NSW 2006, Australia}

\date{\today}

\begin{abstract}
 A new approach to single-qubit operations using exchange interactions of single electrons in gate-defined quantum dots has recently been demonstrated: the resonant exchange qubit. We show that two-qubit operations, specifically the CPHASE gate, can be performed between resonant exchange qubits very straightforwardly, using a single exchange  pulse. This is in marked contrast to the best known protocols for exchange qubits where such a gate requires many pulses so that leakage processes arising from the exchange interaction can be overcome. For resonant exchange qubits a simple two-qubit gate is possible because in this mode of operation energy conservation suppresses leakage. 
 \end{abstract}

\maketitle


Since the seminal work of Loss and Divincenzo~\cite{loss1998a}, the spins of individual electrons trapped in gate-defined quantum dots have been a very promising architecture for quantum computing~\cite{petta2005a,koppens2006a,nowack2007a,pioroladriere2008a,barthel2009a,foletti2009a,shulman2012a,laird2010a,gaudreau2012a,medford2013b}. Various modes of operation have been implemented that pose different experimental challenges. 
An early theoretical advance was the realisation that if a single qubit is encoded in three electron spins then universal operations can be performed using modulated exchange coupling alone~\cite{divincenzo2000a}. 
This in principle removes the need for individually addressable electron spin resonance or large magnetic field gradients. There have been a series of experiments designed to pursue this avenue for quantum computing, despite the seeming inefficiency of using three electron spins to encode a single qubit~\cite{laird2010a,gaudreau2012a,medford2013b}.

A key experimental challenge of the exchange qubit of~\cite{divincenzo2000a} is the complicated pulse sequences that are used to implement two-qubit gates, and these have yet to be experimentally demonstrated. The pulse sequences are required to reverse spin flip transitions induced by exchange coupling between qubits. 
DiVincenzo {\it et al} performed a numerical search to find an explicit pulse sequence to implement a CNOT gate between the two qubits that required 19 exchange pulses in 13 time-steps~\cite{divincenzo2000a}. This pulse sequence was subsequently shown to correspond to an exact analytic solution~\cite{kawano2005a}. A variant of this scheme, known as the decoherence free subsystem, has protection against uniform magnetic fields in all directions and recent results have found pulse sequences for this encoding that are not significantly more complicated, requiring 22 pulses in 13 time-steps~\cite{fong2011a}.  

Recently an alternative approach to exchange-only qubits has arisen, the resonant exchange qubit~\cite{taylor2013a,medford2013a}. As in the exchange qubit of~\cite{divincenzo2000a} a single qubit is encoded in the spin state of three electrons, each of which is trapped in an individual quantum dot. The qubit is operated with significant exchange interaction between the three dots and universal single-qubit operations are perfotmed by rf gate pulses that modulate this exchange interaction. It has been shown both theoretically~\cite{taylor2013a} and experimentally~\cite{medford2013a} that this qubit has several advantages, including first-order insensitivity to charge fluctuations and reduced leakage error due to nuclear field fluctuations. This is a very promising demonstration of a qubit provided that one can couple them efficiently with low leakage. Taylor {\it et al.} show that this can be done through dipole interactions between the qubits~\cite{taylor2013a}, while we suggest the alternative of using exchange coupling between nearby qubits.

In this paper we demonstrate that a further advantage of the resonant exchange qubit is that exchange-based two-qubit gates can be performed very simply. The key idea is that the leakage processes that arise from spin flips do not conserve energy in the resonant exchange qubit.  During an exchange pulse between two qubits these leakage processes are very strongly suppressed and two-qubit gates such as CPHASE can be performed in a single pulse in a single time-step. Moreover, leakage can be made arbitrarily small by reducing the strength of the exchange coupling relative to the resonant qubit's energy splitting. This mechanism contrasts with the conventional approach to exchange-only qubits where the leakage processes are reversed using long pulse sequences. The physics that leads to non-trivial two-qubit gates with low leakage using exchange coupling has also been proposed as the mechanism for two-qubit gates in the context of spin cluster qubits~\cite{meier2003a,meier2003b} and some of our analysis is similar to the discussion in those references. Moreover, the idea of energetically suppressing spin flips for quantum gates in quantum dot qubits specifically has previously arisen in the slightly different context of magnetic field gradients and one- or two-spin qubits~\cite{meunier2011a,klinovaja2012a,wardrop2013a}. 
Exchange-based entangling gates between two resonant exchange qubits would appear to be feasible in the near future and in the rest of the paper we investigate this scheme in more detail.

\begin{table}[htdp]
\begin{center}
\begin{tabular}{|l|l|r|l|}
\hline
 {\rm State} & $S$ & $m_z$ & {\rm Energy}    \\
 \hline
$|Q_{-\frac{3}{2}}\rangle$ & $ \frac{3}{2}$ & $-\frac{3}{2}$& $3B/2$ \\
 $ \left|Q_-\right\rangle$&  $\frac{3}{2}$&$ -\frac{1}{2}$ &   $B/2$  \\
 $\left|1_-\right\rangle$&  $\frac{1}{2}$ & $-\frac{1}{2}$ &  $B/2-J_z/2$   \\    
 $\left|0_-\right\rangle$&  $\frac{1}{2} $& $-\frac{1}{2}$ &  $B/2-3J_z/2$  \\
 $|Q\rangle$& $\frac{3}{2}$ &$ \frac{1}{2}$ &   $ -B/2$  \\
$ |1\rangle$&  $\frac{1}{2}$ & $\frac{1}{2} $ & $-B/2 -J_z/2$  \\    
 $|0\rangle$&$ \frac{1}{2}$ & $\frac{1}{2}  $ & $-B/2-3J_z/2$ \\ 
  $ |Q_{\frac{3}{2}}\rangle$& $\frac{3}{2} $& $\frac{3}{2} $& $-3B/2$ \\ 
   \hline
 \end{tabular}
\end{center}
\caption{Energy eigenstates of the triple-dot system that constitutes a single resonant exchange qubit. $S$ is the total angular momentum quantum number of the three electron spins and $m_z$ is the $z$-component of the total angular momentum, $J_z$ is the exchange splitting for the qubit and $B$ is the Zeeman splitting energy for a single spin.}
  \label{tb:onequbit}
\end{table}%

A single resonant exchange qubit is a triple-dot system operated deep in the $(1,1,1)$ charge state with one electron in each dot. A large uniform magnetic field is applied in the $z$-direction such that the 8 electron spin states are Zeeman split according to the $z$-component of their total angular momentum $m_z$. Tunnelling between adjacent dots is tuned by applied gate voltages resulting in two exchange splittings $J_l$, coupling dot 1 to dot 2, and $J_r$, coupling dot 2 to dot 3. As explained in~\cite{taylor2013a} and in~\cite{medford2013a}, whose notation we largely follow, the qubit operating point has $J_l=J_r$. The qubit states $|0\rangle = \frac{1}{\sqrt{6}}(|\uparrow \uparrow \downarrow\rangle+| \downarrow\uparrow \uparrow\rangle-2| \uparrow\downarrow \uparrow\rangle) $ and $|1\rangle = \frac{1}{\sqrt{2}}(|\uparrow \uparrow \downarrow\rangle-| \downarrow\uparrow \uparrow\rangle )$ are eigenstates of the Hamiltonian describing the exchange interaction between electrons in each dot. They have $m_z=1/2$ and $S=1/2$, where $S$ is the total angular momentum quantum number. The qubit states are split in energy by $J_z=(J_l+J_r)/2$, with $|0\rangle$ being the lower energy state. In~\cite{medford2013a} $J_z/h$ is in the range 0.2--2 GHz.

A third eigenstate $|Q\rangle = \frac{1}{\sqrt{3}}(|\uparrow \uparrow \downarrow\rangle+| \downarrow\uparrow \uparrow\rangle+| \uparrow\downarrow \uparrow\rangle  )$ has $m_z=1/2$ and therefore the same Zeeman energy as the two qubit states, but $S=3/2$. 
The five remaining spin states have different Zeeman splittings and are given by $|Q_{\frac{3}{2}}\rangle=|\uparrow\uparrow\uparrow\rangle$ and $|Q_{-\frac{3}{2}}\rangle,|Q_{-}\rangle,|0_-\rangle,|1_-\rangle$ which are obtained from $|Q_{\frac{3}{2}}\rangle,|Q\rangle,|0\rangle$ and $|1\rangle$ respectively, by flipping all spins. The energies of these states of the single qubit system are indicated in Table~\ref{tb:onequbit}. (In the rest of this paper we will take $\hbar=1$ so that the entries in the table can be regarded as either energies or angular frequencies.) Single qubit operations are performed using oscillatory exchange pulses~\cite{taylor2013a,medford2013a}.

\begin{figure}[htbp]
\begin{center}
\includegraphics[height=4cm]{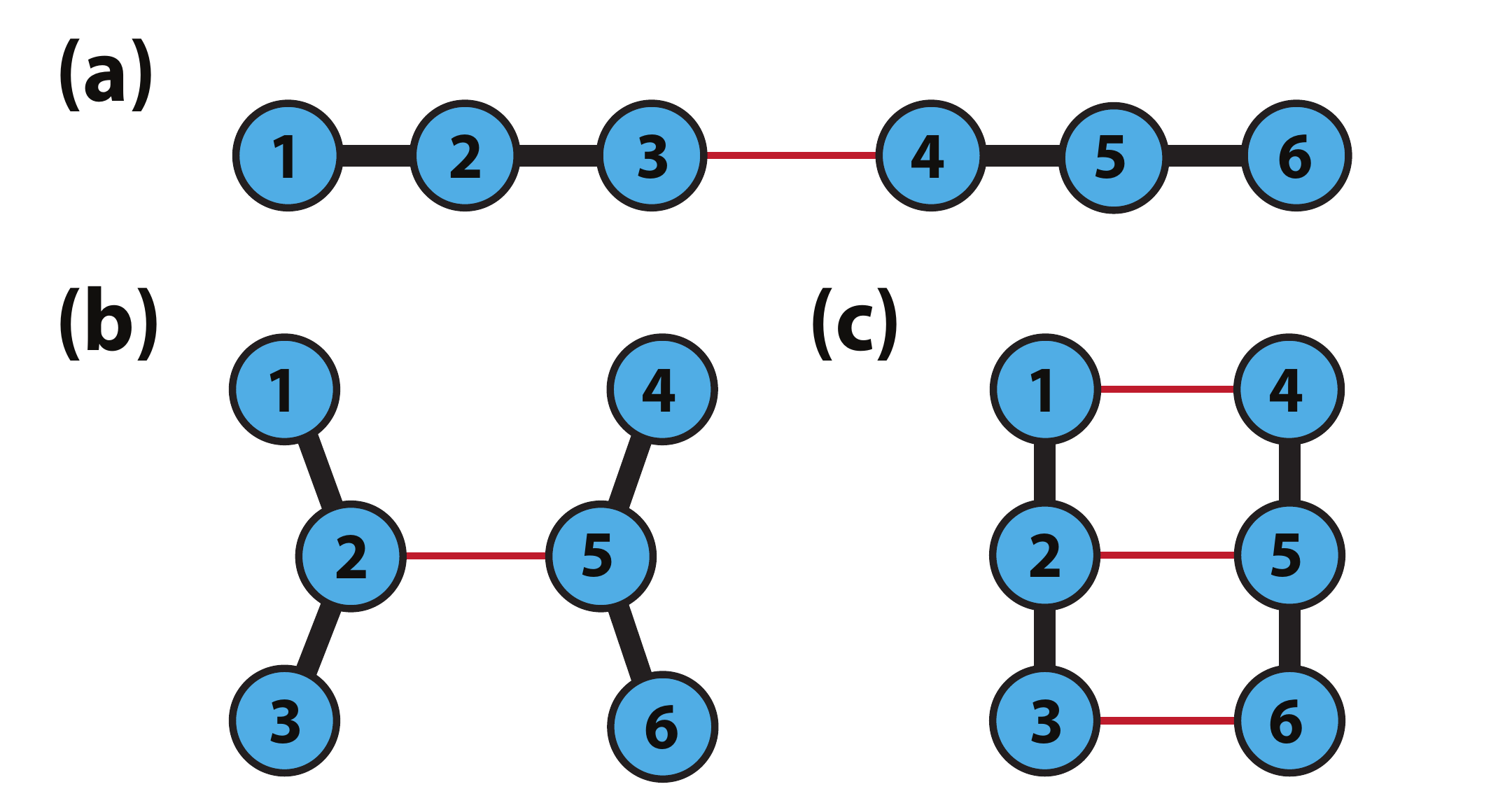}
\caption{Diagram of three geometries for coupling resonant exchange qubits. Solid lines indicate the large exchange coupling $J_z$ within a single qubit. Thin lines indicate the weaker exchange coupling $J_c$ that is turned on between qubits to produce two-qubit gates. a) Linear geometry, b) Butterfly geometry, c) Rectangular geometry.}
\label{fig:schematic}
\end{center}
\end{figure}
%

We now consider a device involving two such qubits, shown schematically in Figure~\ref{fig:schematic}. The first qubit involves dots 1,2 and 3, while the second qubit will involve dots 4,5 and 6. We will label the first (second) qubit $A$ ($B$) and use $J_{zA}$ ($J_{zB}$) to indicate the exchange splitting between the qubit states. We will not assume that $J_{zA}=J_{zB}$. 

Consider the system with no coupling between the qubits, the eigenstates are just the $8^2=64$ tensor products of the states in Table~\ref{tb:onequbit}. The four possible qubit states are in the $m_z=1$ subspace of the six-spin system. This subspace has 15 states, so there are 11 leakage states with the same Zeeman energy that can in principle be coupled by inter-qubit exchange pulses. Table~\ref{tb:twoqubits} indicates these eigenstates and energies. Whatever exchange coupling is turned on between the qubits, it will also conserve total angular momentum. Six linear combinations of these leakage states have total angular momentum $S=2$ or $3$ and so cannot couple to the qubit states which have $S=1$. Moreover exchange couplings will not mix states that have different values of $m_z$. As a result we may focus attention on the block of states with $m_z=1$ and neglect other states from now on.

\begin{table}[htdp]
\begin{center}
\begin{tabular}{|l|l|}
\hline
 {\rm State} & {\rm Energy}+B    \\
 \hline
$|Q,Q\rangle,|Q_{\frac{3}{2}},Q_-\rangle,|Q_-,Q_{\frac{3}{2}}\rangle$ & $0$ \\
$|1,Q\rangle,|1_-,Q_{\frac{3}{2}}\rangle$ &$ -J_{zA}/2$ \\
$|Q,1\rangle,|Q_{\frac{3}{2}},1_-\rangle$ &$ -J_{zB}/2 $\\
$|1,1\rangle$ & $-(J_{zA}+J_{zB})/2$ \\
$|0,Q\rangle,|0_-,Q_{\frac{3}{2}}\rangle$  & $-3J_{zA}/2$\\
$|Q,0\rangle,|Q_{\frac{3}{2}},0_-\rangle$& $-3J_{zB}/2$ \\
$|0,1\rangle$ &$ -(3J_{zA}+J_{zB})/2$\\
$|1,0\rangle$ & $-(J_{zA}+3J_{zB})/2$\\
$|0,0\rangle$ & $-3(J_{zA}+J_{zB})/2$\\
   \hline
\end{tabular}
\end{center}
\caption{Energy level diagram for two resonant exchange qubits in the subspace with $m_z=1$. $J_{zI}$ is the exchange splitting of qubit $I=A,B$. Without loss of generality we have assumed that $J_{zB}\geq J_{zA}$. Note that no qubit state has the same energy as any leakage state. All energies are shifted by the Zeeman energy which is $-B$ for all the states in the table.}
\label{tb:twoqubits}
\end{table}%

The most important feature of Table~\ref{tb:twoqubits} is that none of the leakage state energies correspond to qubit state energies. So, for example, when $J_{zA}=J_{zB}=J_z$ each qubit state is detuned from all leakage states by an energy of at least $J_z/2$. This means that spin flips that result from exchange coupling between the qubits will be energetically unfavourable and leakage rates will be reduced so long as the exchange coupling between qubits remains smaller than the resonant qubit's energy splitting $J_z$. This will mean that two-qubit gates will take a time long compared to  $1/J_z$. However, the single qubit operations for resonant exchange qubits must already run slowly compared to this timescale, so this is not a restriction in practice. We will show below how to perform non-trivial two-qubit gates in a time of order $25(2\pi/ J_z)$ for which leakage error is not expected to be the dominant source of error. For the parameters of~\cite{medford2013a} this time is around $65$ ns and compares favourably with the time taken for single-qubit operations.

We will now explain several approaches to implement specific two-qubit gates using exchange coupling between the two sets of three dots.
The qubits are coupled by inter-qubit Heisenberg exchange coupling $H_{ij}(J_c)=J_c(\sigma_{xi}\sigma_{xj}+\sigma_{yi}\sigma_{yj}+\sigma_{zi}\sigma_{zj}-I)/4$ of the ith dot with the jth dot, where $i\in \{1,2,3\}$ and $j\in \{4,5,6\}$ and $J_c$ is the corresponding exchange energy and is controllable by adjusting some combination of gate voltages. (One could also turn on several such couplings.) 

We will consider three different geometries, as indicated in Figure~\ref{fig:schematic}; the linear, butterfly and rectangular geometries. In the linear arrangement one turns on an exchange interaction $H_{34}$ between the third and fourth dot. In the butterfly arrangement one couples the second and fifth dots ($H_{25}$). Finally in the rectangular arrangement one can couple three sets of of neighbouring pairs of dots through $H_{14}+H_{25}+H_{36}$.  These geometries have different advantages and disadvantages in terms of the simplicity with with which certain gates can be obtained and their susceptibility to noise. However, the choice of geometry will also depend on  which arrangement of dots can most easily be tuned into the $(1,1,1,1,1,1)$ charge configuration and which geometry best allows for coupling large arrays of qubits. 

As noted above leakage processes will be suppressed when $J_c\ll J_z$ so we first analyse these couplings in lowest order perturbation theory. This requires us to calculate the level shifts of the two non-degenerate qubit states $|00\rangle$ and $|11\rangle$, and to find matrix elements of the coupling $H_{ij}$ on the subspace spanned by the two possibly degenerate states $|01\rangle,|10\rangle$. To write the answer explicitly we restrict attention to the qubit subspace and define logical Pauli operators such that $\sigma_{zA}|0\rangle_A=|0\rangle_A$ and  $\sigma_{zA}|1\rangle_A=-|1\rangle_A$, $\sigma_{xA}|0\rangle_A=|1\rangle_A$, etc, and likewise for qubit $B$. Then the Hamiltonian $H_0$ describing the two uncoupled resonant exchange qubits can be read off from Table II and is as follows
\begin{equation}
H_0=-(J_{zA}+J_{zB}) - J_{zA}\sigma_{zA}/2-J_{zB}\sigma_{zB}/2.
\end{equation}
The overall level shift given by the first term is not important so long as we can ignore the leakage states.
For each of the dot geometries the coupling of the two qubits in lowest order perturbation theory can be summarised by the following effective Hamiltonian 
\begin{eqnarray}
H_c&=&\delta J_0 +  \delta J_{z}(\sigma_{zA}+\sigma_{zB})/2 \nonumber \\ &&+ J_{zz} \sigma_{zA}\sigma_{zB} + J_{\perp}(\sigma_{xA}\sigma_{xA}+\sigma_{yA}\sigma_{yA}). 
\label{eq:int}
\end{eqnarray}
where the coefficients $\delta J_0,\delta J_z,J_{zz},J_\perp$ are each proportional to $J_c$, the applied exchange coupling, and are straightforward to calculate. Table~\ref{tb:hamparam} gives explicit values for $\delta J_z,J_{zz},J_\perp$ in each of the three geometries at lowest order in $J_c$. The first term describes a level shift of the qubit states relative to the leakage states. The second term describes a shift of the qubit level splittings $J_{zA},J_{zB}$. This will generate single qubit unitaries that can be corrected if necessary.  The final two terms couple the two qubits and can be used to implement two-qubit gates. Beyond lowest order in perturbation theory the structure of equation~(\ref{eq:int}) will be preserved but the parameters have a more complicated dependence on $J_c$. Note that, similar analyses of Heisenberg interactions of clusters of spins appear in many places in the literature, for example~\cite{meier2003a,meier2003b,srinivasa2009a}.

\begin{table}[htdp]
\begin{center}
\begin{tabular}{|l|lll|}
\hline
 {\rm Geometry} & $\delta J_z/J_c$ & $J_{zz}/J_c$ & $J_\perp/J_c$   \\
 \hline
Linear & 1/36 & 1/36 & -1/24 \\
Butterfly & -1/18 &  1/9 & 0 \\
Rectangular & 0  & 1/6 & 1/12 \\
   \hline
\end{tabular}
\end{center}
\caption{Qubit coupling parameters arising from lowest order perturbation theory in each of the three geometries of Figure~\ref{fig:schematic}. Calculations assume that $J_{zA}\simeq J_{zB}$. When $|J_{zB}-J_{zA}|\gg J_c$ we find $J_\perp=0$ for all geometries, but the other entries in the table are unaffected. }
\label{tb:hamparam}
\end{table}%

We will now describe how to use the interaction of equation (\ref{eq:int}) to implement specific gates. We will largely restrict our attention to CPHASE gates, although other choices are possible. CPHASE is the unitary ${\rm diag}(1,1,1,-1)$. In general we will directly obtain a gate that is equivalent to CPHASE up to local unitaries, (CNOT is one such gate), but since universal single qubit operations are known for the resonant exchange qubit this is enough to provide a universal gate set. The first general approach, the DC scheme, involves slowly turning on and off a small $J_c$. With an interaction time of order $1/J_c$ we obtain a CPHASE gate. We will also consider AC schemes where $J_c(t)$ is modulated at high frequency. 
 
 The butterfly configuration provides an especially simple CPHASE due to the enhanced symmetry in this case. For this configuration we find $J_{zz}=J_c/9$ and $J_\perp=0$ so the two qubits have an effective Ising interaction and an exchange pulse $J_c(t)$ having area $\phi=\int J_c(t)dt= 9\pi/4$ results in a unitary $U(\phi)$ that is equivalent under local unitaries to CPHASE. 

For the linear and rectangular geometries the most straightforward way to obtain the CPHASE gate requires that the two qubits have rather different exchange splittings $| J_{zB}-J_{zA}|\gg J_c$, in which case the qubit states $|01\rangle$ and $|10\rangle$ are no longer degenerate and so $J_\perp=0$. Again a CPHASE results from a single pulse $J_c(t)$ having the correct area such that $\int J_{zz}(t)dt=\pi/4$. If it is preferable to operate in the regime where $J_{zA}=J_{zB}$ then achieving a CPHASE gate will require some single-qubit gates to echo out the effect of the $J_\perp$ term in equation~(\ref{eq:int}). Methods to find explicit pulse sequences exist, see for example~\cite{zhang2002a} and references therein. One simple example is given in the Supplementary Material.

There are three main sources of errors for this gate: charge noise (which leads to noise on $J_c$), magnetic field noise, and leakage. Whereas single qubit operations for the resonant exchange qubit are first-order insensitive to charge noise, it is clear from Equation~(\ref{eq:int}) that fluctuations in $J_c$ will lead to correlated dephasing of the individual qubits through the $\delta J_z\propto J_c$ term as well as to fluctuations in the pulse area $\phi$, reducing the fidelity of the gate. Specifically, low-frequency charge fluctuations will be significant as in the singlet-triplet qubit~\cite{shulman2012a}. Note that since the rectangular geometry has $\delta J_z=0$ it may perform better with regard to charge noise than the other geometries. Nevertheless, experimental experience in two- and three-dot qubit devices~\cite{foletti2009a,shulman2012a,medford2013b} shows that high fidelity exchange pulses of the required length can be carried out in practice. Like single qubit operations of the resonant exchange qubit, the two-qubit gate will be affected by  piezoelectric coupling to phonons~\cite{taylor2013a} and magnetic field fluctuations, specifically to low frequency fluctuations of the z-component of the Overhauser field gradient~\cite{medford2013a}. These lead to independent damping and dephasing of the two qubits and are largely unaffected by the operation of the two-qubit gate. Finally, although the lowest order perturbation theory analysis given above does not have any leakage, this will occur in practice and the exact behaviour will depend on the pulse-shape chosen for $J_c(t)$. Sufficiently smooth pulses with sufficiently small values of $dJ_c/dt$ will greatly suppress leakage relative to a square pulse. 

The error in the gate due to leakage is unlike the other sources of error in that it may be reduced by performing the gate over a longer time, using a smaller value of $J_c$. For example, we can choose to measure the leakage error by maximising the probability of being in a leakage state after the gate over all input qubit states to find the maximum leakage probability $p_L$.  If there were no other source of error the worst case fidelity of the gate would then be $F_L=1-p_L$ (This is a lower bound on the gate fidelity.)  Simple numerical calculations for the butterfly geometry, which has the lowest leakage error of the three geometries, show that when $J_c \leq 0.15 J_z$ the worst case leakage probability $p_L$ is less than $1\%$, even with a square pulse. With $J_z=2\pi\times0.36$ GHz as in~\cite{medford2013a} this leads to a gate time $t_{G}= 9\pi/4J_c \simeq 21$ ns. This compares favourably with the time required for single qubit operations in~\cite{medford2013a}. Leakage errors reduce rapidly as the gate time is increased. So for square pulses with $J_c\leq 0.05 J_z$ having a gate time of around $63$ ns we already find from numerics that the leakage error is below $0.1\%$ .  This shows that with practical gate times leakage error  may be reduced to the point where it is expected to be much less significant than other sources of error. The Supplementary Material has further discussion of leakage.

We briefly consider an AC scheme, where additional control can be obtained by choosing an rf modulation of the the exchange coupling $J_c$. For simplicity, we will consider the rectangular geometry only and choose the two qubits to have different exchange splittings so that $\Delta J_z=J_{zB}-J_{zA}$ is larger than the coupling $J_c$. We can then choose $J_c(t)=J_{c0}(t)+J_{c\Delta}(t)\cos (\Delta J_z t)$ where $J_{c0}$ and $J_{c\Delta}$ are both slowly varying functions of time. Note that we must have $|J_{c\Delta}|<J_{c0}$ since the exchange coupling is always positive. Substituting this into Equation~(\ref{eq:int}) and making the usual rotating-wave approximation, which is valid when $J_{c0},|J_{c\Delta}|\ll \Delta J_z$, we find the following interaction picture Hamiltonian
\begin{eqnarray}
H_{\rm int}&=&\frac{J_{c0}}{6} \sigma_{zA}\sigma_{zA} + \frac{J_{c\Delta}}{24}(\sigma_{xA}\sigma_{xB}+\sigma_{yA}\sigma_{yB}). 
\label{eq:interaction}
\end{eqnarray}
Since $J_{c0}$ and $J_{c\Delta}$ can be independently controlled this interaction allows greater flexibility than in the DC scheme where the parameters in equation (\ref{eq:int}) are all proportional to a single control parameter. So this interaction results in a CPHASE gate in a single pulse of area $\phi=6\pi/4$ with  $J_{c\Delta}=0$ and a SWAP gate in two pulses by choosing $J_{c\Delta}=J_{c0}$ and $\phi=6\pi$ followed by a CPHASE gate with $J_{c\Delta}=0$. 

Since we cannot set $J_{c0}=0$ without also having $J_{c\Delta}=0$, gates based on this scheme remain sensitive to low frequency charge noise (fluctuations in $J_{c0}$) as well as high frequency noise (fluctuations of $J_{c\Delta}$). We are not aware of an exchange-based two-qubit coupling for the resonant exchange qubit that is intrinsically insensitive to noise in $J_c$. However, in the AC scheme echo pulses are available that achieve insensitivity to such low frequency charge fluctuations, an example is given in the Supplementary Material.

In summary we have described a number of protocols for performing two-qubit gates on the recently proposed resonant exchange qubit. For the most convenient parameters and device geometries we have shown how to implement CPHASE  in a single exchange pulse. We leave quantitative estimation of gate fidelities in the presence of realistic noise to future work~\cite{wardrop2013a}.

{\it Acknowledgements:} We would especially like to acknowledge Charles Marcus for bringing this problem to ACD's attention and for good advice and strong encouragement throughout the project. We would also like to acknowledge conversations with James Medford, Stephen Bartlett, and David Reilly, and also Xanthe Croot for help with Figure 1. Research was supported by the Office of the Director of National Intelligence, Intelligence Advanced Research Projects Activity (IARPA), through the Army Research Office grant W911NF-12-1-0354 and by the ARC via the Centre of Excellence in Engineered Quantum Systems (EQuS), project number CE110001013. 



\onecolumngrid
\section{Supplementary information}

\subsection{Echo Pulse Sequences}
In the main text we claimed that the model for two-qubit interaction in the DC scheme given by
\begin{eqnarray}
H_c&=&\delta J_0 +  \delta J_{z}(\sigma_{zA}+\sigma_{zB})/2 \nonumber \\ &&+ J_{zz} \sigma_{zA}\sigma_{zB} + J_{\perp}(\sigma_{xA}\sigma_{xA}+\sigma_{yA}\sigma_{yA}),
\label{eq:intsm}
\end{eqnarray}
allowed for echo pulses to produce CPHASE gates even when $J_\perp\neq 0$.

The simplest example is in the rectangular geometry where $\delta J_z=0$ and all terms in the Hamiltonian commute. If $J_\perp$ were zero we could implement CPHASE with a pulse having $\phi_0=\int J_c(t)dt = 6\pi/4$. The $J_\perp$ term in equation (\ref{eq:intsm}) may be echoed out by breaking the $J_c$ pulse into equal two pieces, implementing the pulse sequence $\sigma_{zA}U(\phi_0/2)\sigma_{zA}U(\phi_0/2)$. The single qubit gates $\sigma_{zA}$ can be implemented using suitable Rabi pulses on the first qubit.  Due to these single qubit echo pulses, the phase acquired as a result of the $J_\perp$ term in the first coupling pulse is unwound in the second pulse. This four stage sequence results in a CPHASE. In the linear arrangement of dots both $\delta J_z$ and $J_\perp$ are non-zero and more complicated pulse sequences are required to obtain a CPHASE gate. 

In the main text we furthermore claimed that  the model for two-qubit interaction in the AC scheme and the rectangular geometry given by
\begin{eqnarray}
H_{\rm int}&=&\frac{J_{c0}}{6} \sigma_{zA}\sigma_{zA} + \frac{J_{c\Delta}}{24}(\sigma_{xA}\sigma_{xB}+\sigma_{yA}\sigma_{yB}), 
\label{eq:interactionsm}
\end{eqnarray}
allowed for echo pulses to produce CPHASE gates while echoing out the, possibly noisy, phase due to $J_{c0}$.

For example, we can choose a pulse with $J_{c\Delta}=J_{c0}$ and pulse area $\tilde{\phi}=\int J_{c\Delta}(t)dt=3\pi$. The gate sequence $\sigma_{xA}U(\tilde{\phi})\sigma_{xA}U(\tilde{\phi})$ results in a gate that is equivalent to CPHASE and is insensitive to low-frequency charge fluctuations. The single-qubit gates $\sigma_{xA}$ can be implemented as a single Rabi $\pi$-pulse on the first qubit. Due to these single qubit echo pulses, the phases acquired as a result of the $J_{c0}$ term and the $\sigma_{yA}\sigma_{yB}$ term in (\ref{eq:interactionsm}) during the first coupling pulse are unwound in the second pulse. The coupling pulses here, $U(\tilde{\phi})$, are each about twice as long as directly implementing CPHASE in the DC scheme. Assuming conservatively that each single qubit pulse takes around the same time as such a direct CPHASE, this gate sequence is about 6 times longer than the direct implementation.

\subsection{Symmetry analysis of coupling}

We present a more detailed discussion of the symmetries of the coupling Hamiltonian in the butterfly geometry in order to justify the statements in the main text about leakage. The extra symmetry in this geometry both acts to reduce leakage errors and enables more detailed analytical calculations. 

As discussed in the text this geometry involves the coupling Hamiltonian $H_{25}(J_c)$.
Notice that this coupling is invariant under swapping the outer dots of either qubit, that is we could swap $1$ and $3$ or $4$ and $6$ and have no effect on the Hamiltonian since it does not act on any of those qubits. These symmetries are also symmetries of the uncoupled two-qubit system. The coupling is also symmetric under swapping the two qubits. This symmetry under swapping qubits is only a symmetry of the full model when $J_{zA}=J_{zB}$.

We should organise the eigenstates of the two-qubit system, as far as possible, according to their parity under these swap operations. If we consider just a single qubit we can notice that the qubit Hamiltonian is invariant under swapping dots 1 and 3 and that $|0\rangle= \frac{1}{\sqrt{6}}(|\uparrow \uparrow \downarrow\rangle+| \downarrow\uparrow \uparrow\rangle-2| \uparrow\downarrow \uparrow\rangle) $ and $|Q\rangle = \frac{1}{\sqrt{3}}(|\uparrow \uparrow \downarrow\rangle+| \downarrow\uparrow \uparrow\rangle+| \uparrow\downarrow \uparrow\rangle  )$ have even parity under this swap while $|1\rangle= \frac{1}{\sqrt{2}}(|\uparrow \uparrow \downarrow\rangle-| \downarrow\uparrow \uparrow\rangle )$ has odd parity. ($SWAP_{13}|1\rangle = -|1\rangle$). Parities of the different tensor products of these states are straightforwardly determined.

As discussed in the text all of the exchange couplings are invariant under global spin rotations and thus we are guaranteed that energy eigenstates can be chosen to have definite total angular momentum quantum numbers. Although this is not true of all the tensor product states discussed in the text it is not difficult to find the correct linear combinations of states with the same energy. We define the following energy eigenstates which all have definite total angular momentum
\begin{eqnarray*}
|E\rangle & =& \sqrt{\frac{3}{5}}|Q,Q\rangle +\frac{1}{\sqrt{5}}\left(\left|Q_{\frac{3}{2}},Q_-\right\rangle+\left|Q_-,Q_{\frac{3}{2}}\right\rangle \right) \\
|E3\rangle &=& \sqrt{\frac{2}{5}}|Q,Q\rangle -\sqrt{\frac{3}{10}}\left(\left|Q_{\frac{3}{2}},Q_-\right\rangle+\left|Q_-,Q_{\frac{3}{2}} \right\rangle\right) \\
|EA\rangle &=& \frac{1}{\sqrt{2}}\left(\left|Q_{\frac{3}{2}},Q_-\right\rangle-\left|Q_-,Q_{\frac{3}{2}} \right\rangle \right) \\
|F\rangle &=& \frac{1}{2} |1,Q\rangle+\frac{\sqrt{3}}{2}\left|1_-,Q_{\frac{3}{2}}\right\rangle \\
|F2\rangle &=&\frac{\sqrt{3}}{2} |1,Q\rangle-\frac{1}{2}\left|1_-,Q_{\frac{3}{2}}\right\rangle \\
|G\rangle &=& \frac{1}{2} |Q,1\rangle+\frac{\sqrt{3}}{2}\left|Q_{\frac{3}{2}},1_-\right\rangle \\
|G2\rangle &=&\frac{\sqrt{3}}{2} |Q,1\rangle-\frac{1}{2}\left|Q_{\frac{3}{2}},1_-\right\rangle \\
|K\rangle &=& \frac{1}{2}|0,Q\rangle+ \frac{\sqrt{3}}{2}\left|0_-,Q_{\frac{3}{2}}\right\rangle  \\
|K2\rangle &=& \frac{\sqrt{3}}{2}|0,Q\rangle- \frac{1}{2}\left|0_-,Q_{\frac{3}{2}}\right\rangle  \\
|L\rangle &=& \frac{1}{2}|Q,0\rangle+ \frac{\sqrt{3}}{2}\left|Q_{\frac{3}{2}},0_-\right\rangle  \\
|L2\rangle &=& \frac{\sqrt{3}}{2}|Q,0\rangle- \frac{1}{2}\left|Q_{\frac{3}{2}},0_-\right\rangle. 
\end{eqnarray*}

These eigenstates have the following symmetry properties
\[
\begin{array}{|c|c|c|c|c|}
\hline
 {\rm State} & {\rm Energy}+B  & {\rm Parity} & {\rm Angular Momentum } \  S & {\rm Swap}    \\
 \hline
|E\rangle & 0 & ++ & 1 & + \\
|E3\rangle & 0 & ++ & 3 & + \\
|EA\rangle & 0 & ++ & 2 & - \\
|F\rangle & -J_{zA}/2 & -+ & 1 &  \\
|F2\rangle & -J_{zA}/2 & -+ & 2 &  \\
|G\rangle & -J_{zB}/2 & +- & 1 &  \\
|G2\rangle & -J_{zB}/2 & +- & 2 &  \\
|1,1\rangle & -(J_{zA}+J_{zB})/2 & -- & 1 & + \\
|M\rangle & -3J_{zA}/2 & ++ & 1 &  \\
|M2\rangle & -3J_{zA}/2 & ++ & 2 &  \\
|N\rangle & -3J_{zB}/2 & ++ & 1 &  \\
|N2\rangle & -3J_{zB}/2 & ++ & 2 &  \\
|0,1\rangle & -(3J_{zA}+J_{zB})/2 & +- & 1 & \\
|1,0\rangle & -(J_{zA}+3J_{zB})/2 & -+ & 1 & \\
|0,0\rangle & -3(J_{zA}+J_{zB})/2 & ++ & 1 & + \\
   \hline
\end{array}
\]
The parity column indicates the parity under swapping the first and third dot and also the fourth and sixth dot as discussed above. Swap indicates the parity of the state under swapping qubits where this is well defined.

\subsection{Leakage}

In this section we justify the statements about leakage made in the main paper. We consider just the butterfly geometry.

The exchange gate coupling Hamiltonian $H_{25}$ can only couple eigenstates with the same total angular momentum and parity quantum numbers. So for example $\langle E3|H_{25}|ij\rangle=0$ where $|ij\rangle$ is any qubit state, because $S=1$ for each qubit state and $|E3\rangle$ has $S=3$.  On the basis of their total angular momentum we conclude that the six states $|E3\rangle,|EA\rangle, |F2\rangle,|G2\rangle,|M2\rangle,|N2\rangle $ do not couple to the qubit states during an exchange pulse. The existence of these uncoupled leakage states was mentioned in the main paper. The five relevant leakage states are $|E\rangle,|F\rangle, |G\rangle |M\rangle |N\rangle$

Each of the qubit states has a different parity. As a result we find for qubit states $\langle ij|H_{25}|kl\rangle=0$ whenever $(i,j)\neq (k,l)$ and we can conclude that when restricted to the qubit subspace $H_{25}$ has no off-diagonal matrix elements. 

The state $|11\rangle$ is the only state with parity $--$ and so it doesn't couple to leakage states at all. Since $|G\rangle$ is the only leakage state with parity $+-$ and $S=1$ it is the only state coupled to $|0,1\rangle$ by the exchange pulse. When $J_{zA}=J_{zB}$ the energy difference between these states is $3J_z/2$. Likewise $|1,0\rangle$ couples to $F$ and the energy difference between these states is also $3J_z/2$. Finally $|0,0\rangle$ couples to the three remaining leakage states $|E\rangle, |M\rangle,|N\rangle$ that are different in energy by $3J_z,3J_z/2$ and $3J_z/2$ respectively.

So in the butterfly geometry the enhanced symmetry means that each computational basis state couples independently to an {\it orthogonal} leakage subspace. As a result we may write a completely general solution for the mapping of the computational basis states under any exchange pulse $J_c(t)$ as follows
\begin{eqnarray*}
|0,0\rangle & \rightarrow & \sqrt{1-p_{00}}e^{-i\varphi_{00}}|00\rangle +   \sqrt{p_{00}}|L_{00}\rangle \\
|0,1\rangle & \rightarrow & \sqrt{1-p_{01}}e^{-i\varphi_{01}}|01\rangle +   \sqrt{p_{01}}|L_{01}\rangle \\
|1,0\rangle & \rightarrow & \sqrt{1-p_{10}}e^{-i\varphi_{10}}|10\rangle +   \sqrt{p_{10}}|L_{10}\rangle \\
|0,0\rangle & \rightarrow &e^{-i\varphi_{11}}|00\rangle  \\
\end{eqnarray*}
Where $|L_{00}\rangle ,|L_{01}\rangle $ and $|L_{10}\rangle $ are orthogonal leakage states. The leakage probability associated with the computational basis state $|ij\rangle$ is $p_{ij}$ and because of the orthogonality of the $|L_{ij}\rangle$ the largest of these is the maximum leakage probability over all possible input two-qubit states. This is the maximum leakage error $p_L$ discussed in the main text. The symmetry of the problem requires that $p_{01}=p_{10}$ and typically we find that $p_{01}> p_{00}$, so usually $p_L=p_{01}$.

\begin{figure}[htbp]
\begin{center}
\includegraphics[width=8.5cm]{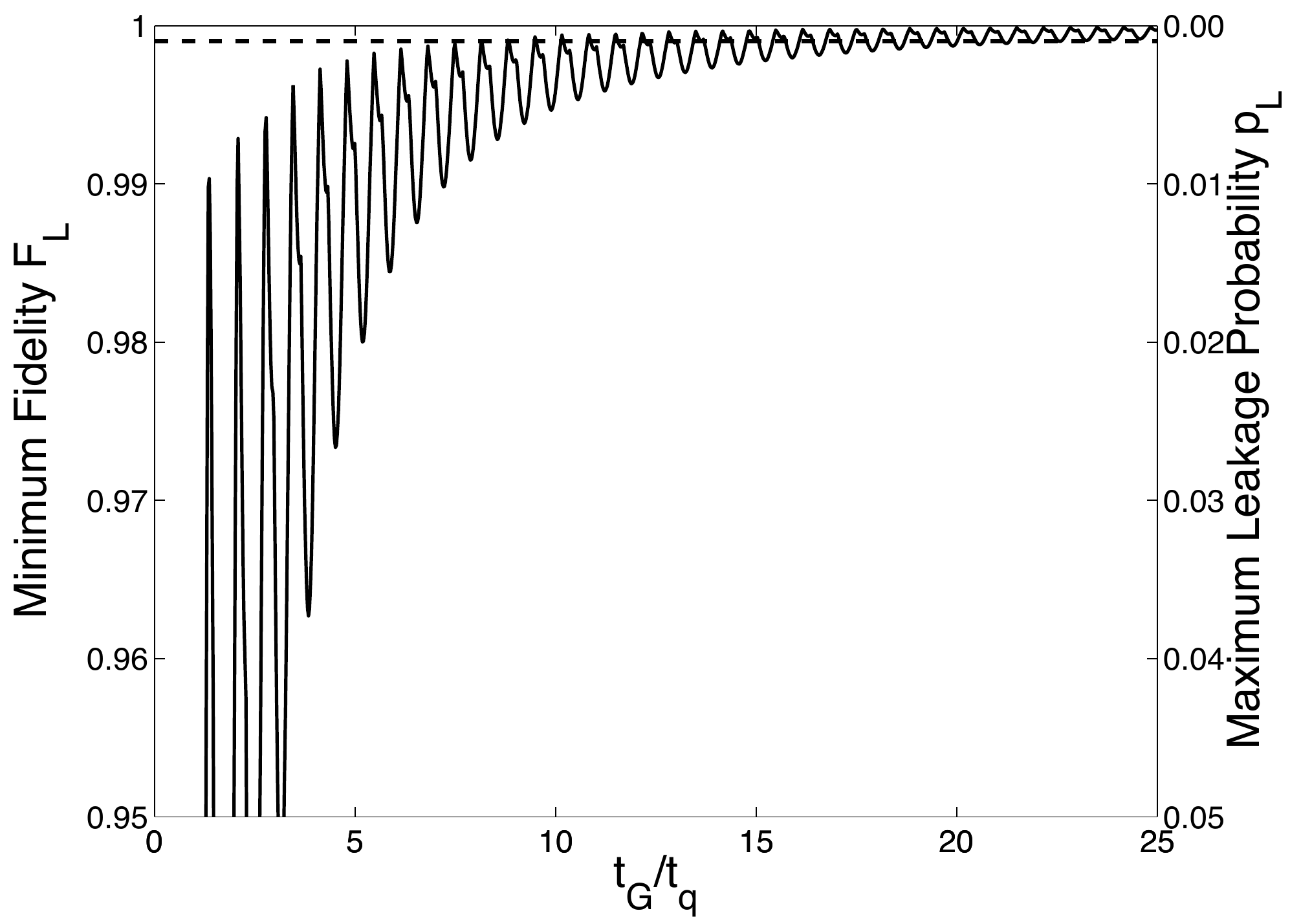}
\caption{Plot of minimum gate fidelity $F_L$ and worst case leakage probability $p_L$ for a two-qubit gate having $\phi=\int J_c(t)=9\pi/4$ and $J_{zA}=J_{zB}$ in the butterfly geometry, using a square pulse of time $t_G$. Time is normalised by $t_q=2\pi/J_z$, the Bohr period of each resonant exchange qubit, which is around $3$ ns for the parameters of~\cite{medford2013a}. The dashed horizontal line indicates a minimum fidelity of $0.999$ or $0.1\%$ error. }
\label{fig:leakage}
\end{center}
\end{figure}

To obtain a gate equivalent to CPHASE up to local unitaries one chooses the pulse area so that $\exp[-i(\varphi_{00}+\varphi_{11}-\varphi_{01}-\varphi_{10})]=-1$. Once again due to the orthogonality of the leakage states, the minimum fidelity of this gate over all possible input two-qubit states is $F_L=1-p_L$ in the absence of other sources of error. These quantities are plotted as a result of numerical integration for a square pulse of fixed area in Figure 1. These results are the basis of the quantitative statements about leakage in the main text.  As noted in the main text fidelity is greatly improved, for the same gate time, by using smooth pulses rather than square pulses~\cite{wardrop2013a}.

\end{document}